\documentclass[]{spie}  %>>> use for US letter paper
\usepackage{amsmath,amsfonts,amssymb}
\usepackage{graphicx}
\usepackage[colorlinks=true, allcolors=blue]{hyperref}

\title{An overview of the mid-infrared spectro-interferometer MATISSE: science, concept, and current status}

\author[a]{A. Matter}
\author[a]{B. Lopez}
\author[a]{P. Antonelli}
\author[b]{M. Lehmitz}
\author[c]{F. Bettonvil}
\author[d]{U. Beckmann}
\author[a]{S. Lagarde}
\author[e]{W. Jaffe}
\author[a]{R. G. Petrov}
\author[a]{P. Berio}
% ...

\affil[a]{Laboratoire Lagrange, Universit\'e C\^ote d'Azur, Observatoire de la C\^ote d'Azur, CNRS, Boulevard de l'Observatoire, CS 34229, 06304 Nice, France.}
\affil[b]{Max-Planck Institute for Astronomy, Heidelberg, Germany.}
\affil[c]{NOVA ASTRON, Dwingeloo, the Netherlands.}
\affil[d]{Max-Planck Institute for Radio astronomy, Bonn, Germany.}
\affil[e]{Leiden Observatory, Leiden University, the Netherlands.}
\affil[f]{European Southern Observatory.}
\affil[g]{Institute of Theoretical Physics and Astrophysics, Kiel University, Germany.}
\affil[h]{Universit\'e de Grenoble Alpes, CNRS, IPAG, Grenoble, France.}
\affil[i]{Lowell Observatory, Flagstaff, USA.}
\affil[j]{MTA Research Centre for Astronomy and Earth Sciences, Konkoly Thege Miklos Astronomical Institute, Budapest, Hungary.}
\affil[k]{NASA/Goddard Space Flight Center, Greenbelt, USA.}
\affil[l]{Sterrenkundig Instituut Anton Pannekoek, University of Amsterdam, the Netherlands.}
\affil[m]{Institut fuer Astrophysik, University of Vienna, Austria.}
\affil[n]{Physikaliches Institut, University of Cologne, Germany.}
\affil[o]{Laboratoire AIM, CEA/DSM-CNRS-Universit\'e Paris Diderot, IRFU/Service d'Astrophysique, CEA/Saclay, Gif-sur-Yvette, France.}

\authorinfo{Further author information: (Send correspondence to A. Matter)\\A. Matter: E-mail: Alexis.Matter@oca.eu, Telephone: +33 (0)4 92 00 19 79}

\begin{document} 
\maketitle

\begin{abstract}
MATISSE is the second-generation mid-infrared spectrograph and imager for the Very Large Telescope Interferometer (VLTI) at Paranal. This new interferometric instrument will allow significant advances by opening new avenues in various fundamental research fields: studying the planet-forming region of disks around young stellar objects, understanding the surface structures and mass loss phenomena affecting evolved stars, and probing the environments of black holes in active galactic nuclei. As a first breakthrough, MATISSE will enlarge the spectral domain of current optical interferometers by offering the {\itshape L} and {\itshape M} bands in addition to the {\itshape N} band. This will open a wide wavelength domain, ranging from 2.8 to 13~$\mu$m, exploring angular scales as small as 3 mas ({\itshape L} band) / 10 mas ({\itshape N} band). As a second breakthrough, MATISSE will allow mid-infrared imaging - closure-phase aperture-synthesis imaging - with up to four Unit Telescopes (UT) or Auxiliary Telescopes (AT) of the VLTI. Moreover, MATISSE will offer a spectral resolution range from $R \sim 30$ to $R \sim 5000$. Here, we present one of the main science objectives, the study of protoplanetary disks, that has driven the instrument design and motivated several VLTI upgrades (GRA4MAT and NAOMI). We introduce the physical concept of MATISSE including a description of the signal on the detectors and an evaluation of the expected performances. We also discuss the current status of the MATISSE instrument, which is entering its testing phase, and the foreseen schedule for the next two years that will lead to the first light at Paranal.
%This document is prepared using LaTeX2e\cite{Lamport94} and shows the desired format and appearance of a manuscript prepared for the Proceedings of the SPIE.\footnote{The basic format was developed in 1995 by Rick Hermann (SPIE) and Ken Hanson (Los Alamos National Lab.).} It contains general formatting instructions and hints about how to use LaTeX.  The LaTeX source file that produced this document, {\ttfamily article.tex} (Version 3.4), provides a template, used in conjunction with {\ttfamily spie.cls} (Version 3.4). These files are available on the Internet at \url{https://www.overleaf.com}.  The font used throughout is the LaTeX default font, Computer Modern Roman, which is equivalent to the Times Roman font available on many systems.  
\end{abstract}


\section{INTRODUCTION}
\label{sec:intro}  % \label{} allows reference to this section
The Multi AperTure mid-Infrared SpectroScopic Experiment (MATISSE) is the mid-infrared spectrograph and imager planned for the ESO/VLTI. This second generation interferometric instrument
will address several f­undamental research topics in astrophysics, including the inner regions of disks around young stars, where planets form and evolve, the surface structure and mass loss of stars at different evolutionary stages, and the environment of central black holes in active galactic nuclei (AGN). MATISSE offers unique interferometric capabilities. This includes the opening of the {\itshape L} and {\itshape M} bands (respectively 2.8-4.0 and 4.6-5.0 $\mu$m) to long-baseline interferometry. An angular resolution down to about 3 milliarcseconds (mas) in {\itshape L} band and various spectral resolutions between $R\sim 30$ and $R\sim 5000$ will be available. The second unique capability will be mid-infrared imaging performed with the four Unit Telescopes (UTs) and Auxiliary Telescopes (ATs). For this reason, MATISSE can be seen as a successor to MIDI (the MID-infrared Interferometric instrument) \cite{2003Ap&SS.286...73L} at the VLTI.
MATISSE will link the near-infrared spectral domain, covered currently at the VLTI by AMBER\cite{2007A&A...464....1P} and soon by the second generation instrument GRAVITY\cite{2011Msngr.143...16E}, with the millimeter domain, where the Atacama Large Milli­meter/Submillimeter Array (ALMA) provides a similar angular resolution. MATISSE will also be complementary to the instruments foreseen for the E-ELT, in particular METIS (the Mid-infrared E-ELT Imager and Spectrograph). While MATISSE will provide an angular resolution higher by a factor $\sim 4$ to 5, METIS will yield a higher sensi­tivity, higher spectral resolution and a broader wavelength coverage.\\
We present some of the main science objectives that have driven the instrument design. We introduce the physical concept behind MATISSE, including a description of the signal on the detectors and an evaluation of the expected performance, and we discuss the current status of the project. The operations concept will be detailed in a future article, which will illustrate the observing templates that operate the
instrument, the data reduction and image reconstruction software.

\section{Genesis of the project}
\label{sec:genesis}
Following the first light of the two-telescope VLTI instrument MIDI in 2002, ideas already came out on a possible upgrade towards an mid-infrared interferometric imager. A first prototype, called APreS-MIDI (Aperture SynthesiS with MIDI), was built at OCA in Nice. This prototype was presented at the ESO VLTI conference in 2005\cite{2008poii.conf.....R}. Following a recommendation by ESO, the MATISSE Consortium formed and initiated a ­conceptual design study for a second generation VLTI instrument. The ­Preliminary Design Review of MATISSE was held in December 2010 in ESO-Garching. Then the Final Design Reviews occurred in September 2011 for cryogenics and optics, and in April 2012 for the whole instrument.
Currently, the instrument is fully integrated and the test phase in the laboratory has started in Nice Observatory. The preliminary acceptance in Europe is
planned for mid-2017, for a first light at Paranal late 2017.

\section{Scientific motivation}
\label{sec:science}
From the very beginning of the project, MATISSE was planned as an interferometric imager for a broad range of astrophysical targets. To achieve this goal, stringent requirements for the instrument were derived from the most challenging
science cases\cite{Lopez13}: the protoplanetary disks around progenitors to
Solar-type stars (T-Tauri stars) and the dusty tori around AGN. The expected and achieved instrument characteristics will also open up other fields: the birth, structure, dynamics and chemistry of massive and evolved stars; the early evolution of Solar System minor bodies; the exozodiacal disks around main-sequence stars; the properties of hot Jupiter-like exoplanets; and
the study of the immediate vicinity of the Galactic Center. In
the following, we present an overview of the main science case that has driven the MATISSE project: the protoplanetary disks. A complete overview of the MATISSE science cases is provided in Wolf et al. (2016, this volume).
\begin{table}[ht]
\caption{Selected spectral signatures accessible with MATISSE.
%Fonts sizes to be used for various parts of the manuscript.  Table captions should be centered above the table.  When the caption is too long to fit on one line, it should be justified to the right and left margins of the body of the text.
} 
\label{tab:signatures}
\begin{center}       
\begin{tabular}{ll} %% this creates two columns
%% |l|l| to left justify each column entry
%% |c|c| to center each column entry
%% use of \rule[]{}{} below opens up each row
\hline
\multicolumn{2}{c}{{\itshape L}\&{\itshape M} bands ($\sim 2.8-5.0$~$\mu$m)} \\
\hline
  {\small H$_2$O (ice)} & 3.14~$\mu$m  \\
  {\small H$_2$O (gas)} & {\small $2.8-4.0 \mu$m}  \\
    {\small H lines (Br-$\alpha$, Pf-$\beta$)} & {\small 4.05, 4.65~$\mu$m}   \\
 {\small PAHs} & {\small 3.3, 3.4~$\mu$m}   \\
  {\small Nanodiamonds} & {\small 3.52~$\mu$m }  \\
 {\small CO fundamental transitions (gas)} & {\small $4.6-4.78 \mu$m}  \\
 {\small CO (ice)} & {\small $4.6-4.7 \mu$m}  \\
\hline
\multicolumn{2}{c}{{\itshape N} band ($\sim 8.0-13.0$~$\mu$m)} \\
\hline
 {\small Amorphous silicates}  &{\small 9.8~$\mu$m}  \\
 {\small Crystalline silicates (olivines, pyroxenes)} &{\small 9.7, 10.6, 11.3, 11.6~$\mu$m}\\
 {\small PAHs}  & {\small 8.6, 11.4, 12.2, 12.8~$\mu$m} \\
 {\small Fine structure lines (e.g., [Ne II])}  &  {\small 10.5, 10.9, 12.8~$\mu$m}\\
\hline 
\end{tabular}
\end{center}
\end{table} 
\subsection{Protoplanetary disks}
The investigation of the potential planet-forming regions around young stars in nearby star-forming regions has been made possible by infrared interferometry. The large-scale ($\sim 10-100$~au) characteristics of the protoplanetary disks could then be compared to the
structure of the inner, au-scale, regions \cite{2004A&A...423..537L,2009A&A...502..367S,2015A&A...581A.107M,2016A&A...586A..11M}. For instance, differences found in dust grain size and crystallinity provided valuable insights into the disk mineralogy and the radial transport of solids \cite{2004Natur.432..479V}, while the temporal variability of the re-emission brightness on scales of a few au sheds light on processes in young eruptive stars \cite{2013A&A...552A..62M}. 
%In the near-infrared, VLT/AMBER was able to resolve the sub-­au- scale gas and dust regions of accretion discs and the launching areas of winds in the continuum and emission lines (e.g.,
%Br γ). Furthermore, the high spectral resolution offered by AMBER enabled the study of the kinematic properties of inner discs and disc wind regions (e.g., Weigelt et al., 2011). Such studies are important to improve our understanding of the fundamental accretion–ejection process.
In combination with other high angular resolution instruments/observatories operating at complementary wavelength ranges (e.g., GRAVITY and ALMA), MATISSE will provide the means to study
the planet-forming regions, from about 0.1 to 10~au, in detail. Specific key topics and questions that MATISSE will tackle include: the complexity of disk structures in the planet-forming zone of circumstellar disks at various stages of their evolution; the reasons for inner-disk clearing in transitional disk; constraints on properties, growth, and sedimentation of dust grains; the tracing structures of accreting protoplanets; the nature of outbursting young stellar objects; the production of micron-sized dust as an outcome of planetesimal collisions and evaporation of exo-comets; the launching region of winds and jets and the disk-outflow connection. We describe hereafter the specific characteristics and capabilities of MATISSE :
\begin{itemize}
\item the imaging capabilities of MATISSE will enable the direct and non-ambiguous detection of au-scale complex disk structures in the potential planet-forming regions. Structures that can be used as tracers for the mechanisms of planet formation.
\item The extension to the {\itshape L} and {\itshape M} bands will allow coverage of the entire spatial range of the inner disk regions, from about 0.1 to 10~au, and investigation of the physical processes at play. While the {\itshape N} band (7.5-14.5 $\mu$m) observations are dominated by
the thermal emission of warm dust ($\sim 300$~K), the {\itshape L}/{\itshape M} band flux is expected to consist of both emission and scattering.
\item MATISSE will offer spectroscopic capabilities with various spectral resolutions in the range of 30 to about 5000. Spatially resolved spectroscopy of various spectral features of solid, ice and gas species will thus be possible (see Table~\ref{tab:signatures}). Such observations will make possible the study and mapping of amorphous and crystalline silicate dust, carbonaceous molecules and water ice, as well as the distribution and kinematics of the hot gas.
\item Repeated observations will allow investigation of the temporal changes suggested from planet formation and planet-disk interaction scenarios.
\end{itemize}

\begin{figure} [t]
   \begin{center}
   \begin{tabular}{c} %% tabular useful for creating an array of images 
   \includegraphics[scale=0.6]{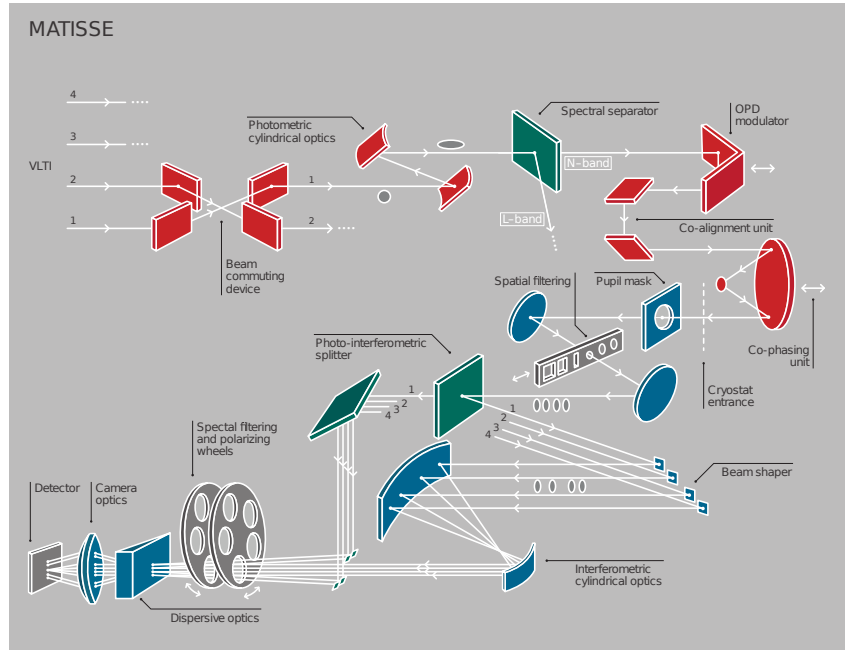}
   \end{tabular}
   \end{center}
   \caption[example] 
%>>>> use \label inside caption to get Fig. number with \ref{}
   { \label{fig:layout} 
Schematic layout of MATISSE. The red parts represent optical elements located on the warm optics table at ambient temperature. The blue parts represent the optical elements of the cold optics bench located in the cryostats. Only one Cold Optics Bench (COB) is shown here.}
   \end{figure}  

\subsection{Upgrades at the VLTI}
\begin{figure} [t]
   \begin{center}
   \begin{tabular}{c} %% tabular useful for creating an array of images 
	\includegraphics[width=150mm,height=70mm]{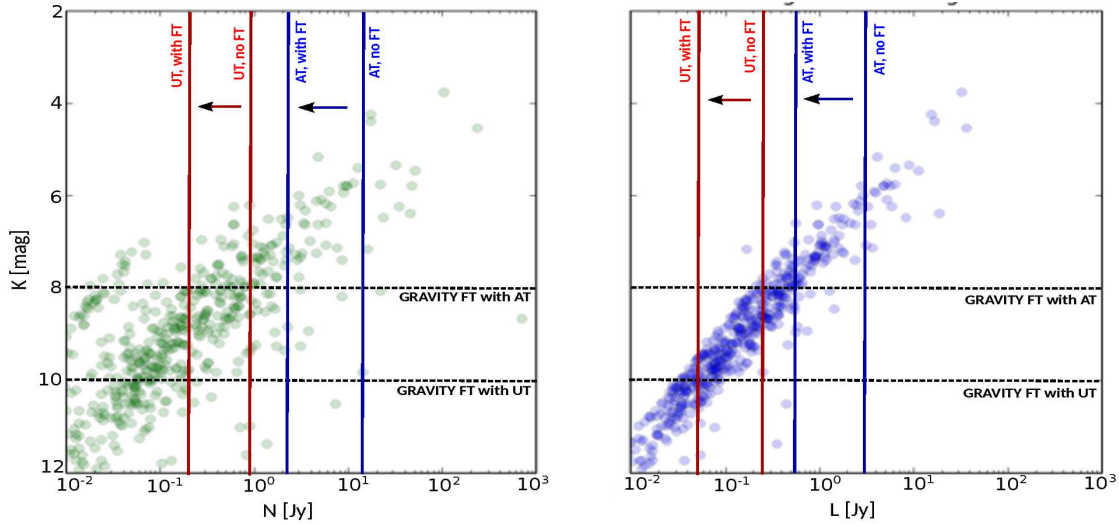}   
   \end{tabular}
   \end{center}
   \caption[example] 
%>>>> use \label inside caption to get Fig. number with \ref{}
   { \label{fig:gra4mat} 
%\textbf{Left: }{\itshape K} and {\itshape L} magnitudes of a panel of low-mass young stars in the Taurus star-forming region. The two vertical lines shows the estimated (ATs) and expected (UTs) {\itshape K-}band sensitivity of the GRAVITY FT, while the horizontal lines show the minimum gain in sensitivity for MATISSE in {\itshape L-}band.
{\itshape K} and {\itshape L} and {\itshape N} fluxes of the 742 pre-main sequence stars (members of the Orion OB association) listed in the Herbig \& Bell catalogue \cite{1988cels.book.....H}. The two horizontal dashed lines shows the estimated (ATs) and expected (UTs) {\itshape K} band sensitivity of the GRAVITY Fringe Tracker (FT), while the horizontal lines show the MATISSE sensitivity without and with the use of GRA4MAT.}
   \end{figure}

Addressing the astrophysical questions described above on protoplanetary disks requires the full exploitation of the potential of MATISSE. In this context, instrumental upgrades at the VLTI will be mandatory. This concerns in particular the implementation of adaptive optics on the ATs, and, most important, the availability of an external fringe tracker for MATISSE. An external fringe tracker will improve the sensitivity, accuracy and spectroscopic capabilities of MATISSE, with a direct impact on the scientific potential of the instrument, in the following ways:
\begin{itemize}
\item Reaching sensitivities below a few tens of mJy in the mid-infrared would allow MATISSE to access a statistically significant number of sources (a few hundred) of different ages and masses, for both model fitting and image reconstruction studies. Probing different age and mass regimes is fundamental to understand the disks evolution scheme down to the faintest sources represented by the solar-mass young stars (T Tauri stars). 
\item High accuracy phase measurements ($\sim 1$~milliradian) will be important for closure-phase imaging in the {\itshape L}, {\itshape M} and {\itshape N} bands, which will especially enable the non-ambiguous detection and study of complex au-scale structures and asymmetries in the inner region of disks.  
\item With the longer integration times allowed by an external fringe tracking device, medium and high spectral resolution interferometry will be made feasible over the full range of the spectral bands of MATISSE.
\end{itemize}
Two upgrades are planned at the VLTI : NAOMI, the Adaptative Optics system for the ATs, and GRA4MAT. The GRA4MAT project will enable the use of the GRAVITY Fringe Tracker (FT) to track fringes in {\itshape K} band while measuring the interferometric observables in {\itshape L}, {\itshape M} and {\itshape N} bands with MATISSE. In this process, the main limiting factor is the {\itshape K} band sensitivity of the GRAVITY FT. This sensitivity has been estimated to $K_{\rm corr}\simeq7$ (correlated magnitude) during the GRAVITY commissioning with the ATs, and is expected to be $K_{\rm corr}\simeq10$ with the UTs.
Compared to these values, we can expect a sensitivity gain of 0.8 to 1 magnitude in the case of GRA4MAT, where all the {\itshape K} band flux will be sent to the fringe tracker. This is not the case for GRAVITY alone, where only 50\% of the flux is available for fringe tracking.
Then, provided fringes can be tracked in {\itshape K} band, we can expect a minimum gain of about 2 magnitudes at the MATISSE wavelengths (see Table \ref{tab:fluxes}). The minimum gain provided by GRA4MAT in the specific case of the observation with MATISSE of low-mass young stars in the Taurus star-forming region, and pre-main sequence stars members of the Orion OB association, is shown in Fig.\ref{fig:gra4mat}. The current plan is to have GRA4MAT commissioned late 2018, at the end of the MATISSE commissioning (see Sect. \ref{sec:status}).   
%However, to make full use of the potential of MATISSE and thus to fully achieve the
%above goals, improvements in the VLTI infrastructure are mandatory. In particu-
%lar, these concern the decrease of the vibration level of the UTs, adaptive optics
%on the ATs, and, most important, the availability of a second generation fringe
%tracker (2GFT) for MATISSE. A 2GFT will improve the sensitivity, accuracy and spectroscopic capability of MATISSE and will thus have a direct impact on the scientific potential of the instrument, in the following ways:
%\begin{itemize}
%\item The sensitivity achievable with a 2GFT is required for the study of AGNs and
%the discs around young stars. For example, with a 2GFT, longer baselines
%can be used to establish the connections between the high surface bright-
%ness inner discs and the asymmetric larger components.
%\item Higher accuracy is important for closure-phase imaging in the {\itshape L}-, {\itshape M}- and {\itshape N}-bands, which provides constraints on the radial and vertical temperature gradient and opacity structure in discs of young stars.
%\item Medium and high spectral resolution interferometry will be made feasible over the full range of the spectral bands.
%\end{itemize}
%Begin the Introduction below the Keywords. The manuscript should not have headers, footers, or page numbers. It should be in a one-column format. References are often noted in the text and cited at the end of the paper.
\begin{figure} [t]
   \begin{center}
   \begin{tabular}{c} %% tabular useful for creating an array of images 
   \includegraphics[scale=0.5]{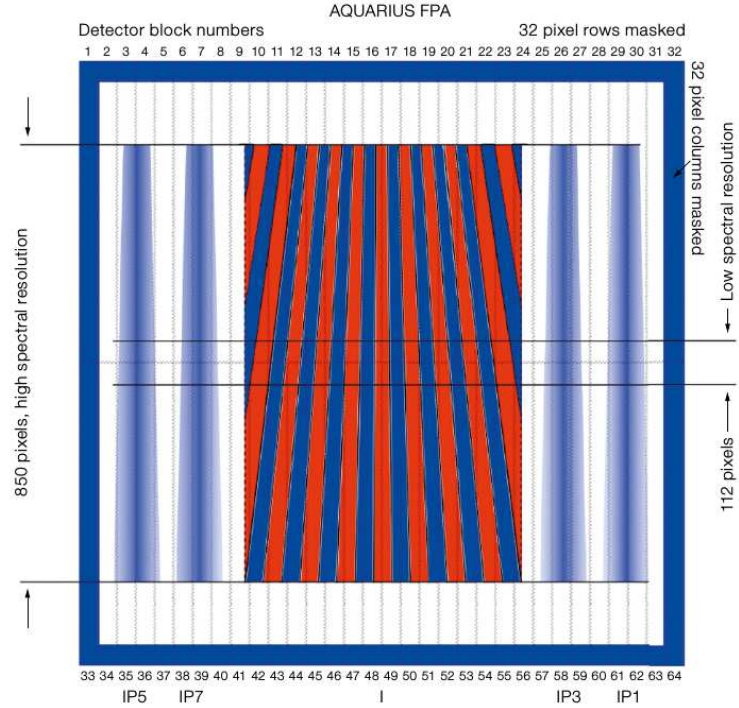}
   \end{tabular}
   \end{center}
   \caption[example] 
%>>>> use \label inside caption to get Fig. number with \ref{}
   { \label{fig:detector} 
Layout of the interferometric pattern (central detector blocks, dispersion direction is vertical) and the four photometric signals on the {\itshape N} band AQUARIUS detector, in SiPhot mode with medium spectral dispersion.}
   \end{figure}
\section{Concept}
MATISSE uses an all-in-one multi-axial beam combination scheme. We concluded that this type of combination is the most suitable for an interferometric instrument with more than two apertures and operating in the mid-infrared. Based on the efficiency of the two-telescope MIDI recombination scheme, a pairwise co-axial concept was initially considered. The advantage of this scheme is the simultaneous delivery of two interferometric signals per baseline, phase shifted by $\pi$ radians. The correlated flux is then obtained by subtracting the two signals. In this way, the thermal background level and its associated temporal fluctuations are directly eliminated, but not the related thermal photon noise. However, in spite of good expected efficiency in terms of signal-to-noise ratio (SNR), this scheme displays a number of issues when extended from two to four telescopes: a possible weakness in the stability of the closure-phase measurements and a high instrumental complexity due to numerous opto-mechanical elements required in cryogenic conditions. These issues led us to consider multiaxial global combination as a more robust and simpler scheme.
\begin{figure}[t]
   \begin{center}
   \begin{tabular}{c} %% tabular useful for creating an array of images 
   \includegraphics[scale=0.65]{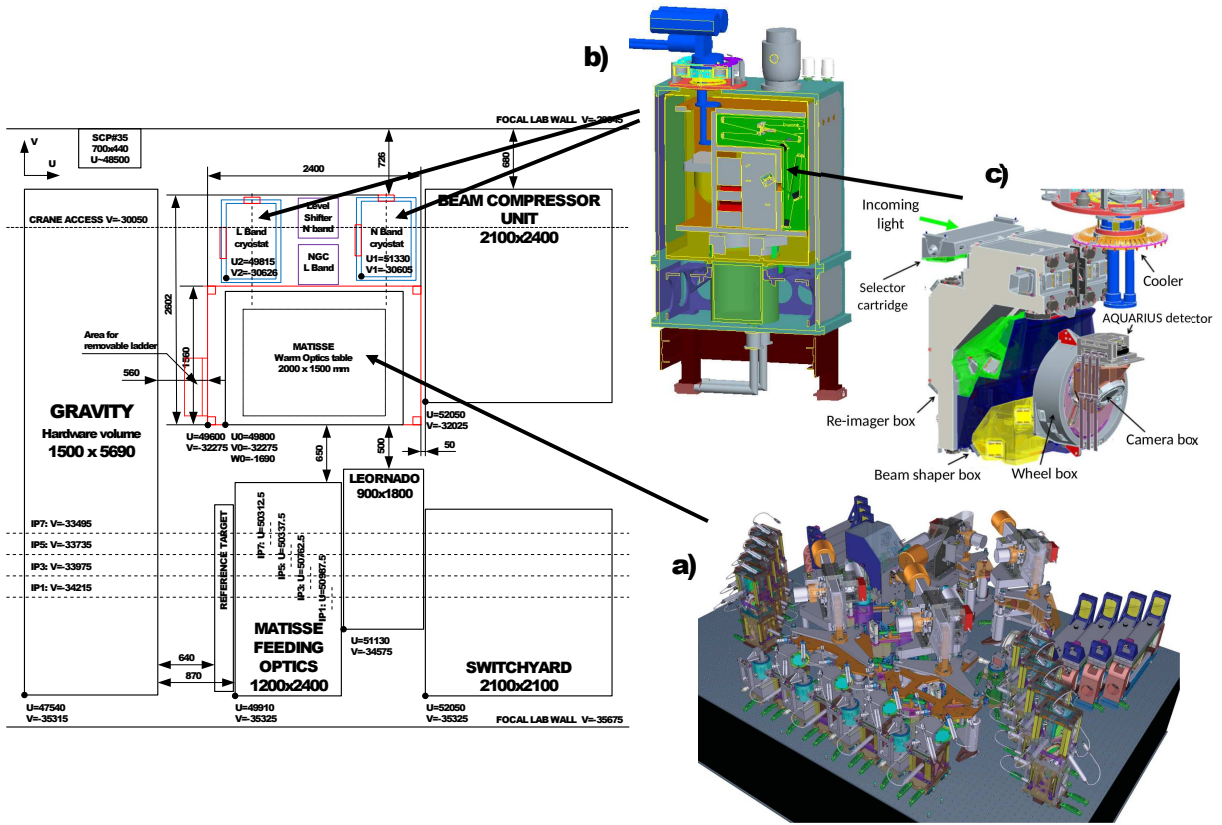}
   \end{tabular}
   \end{center}
   \caption[example] 
%>>>> use \label inside caption to get Fig. number with \ref{}
   { \label{fig:matissevlti} 
The future location of MATISSE in the VLTI is sketched, with a view (from above) of the warm optics table and the two cryostats. a) view of the warm optics table with its optical components; b) One of the two MATISSE cryostats, which holds the cold optics bench and the detector; c) View of one of the two cold optics benches.}
   \end{figure}

In the multi-axial beam combination scheme of MATISSE (see Fig.\ref{fig:layout}), the four beams are combined simultaneously onto the detector on an area called the interferometric channel, while the four individual photometric signals are imaged individually on each side. MATISSE will observe in three bands simultaneously: {\itshape L}, {\itshape M} and {\itshape N}. The interferometric pattern and the photometric signals are spectrally dispersed using grisms: spectral resolutions of 30 and 220 are provided in {\itshape N} band and four resolutions in {\itshape L} and {\itshape M} bands of 30, 500, 1000 and 3500-5000. The spatial extent of the interferometric pattern is larger than the photometric channels in order to optimize the sampling of the six different spatial fringe periods. The beams are combined by the camera optics. At this plane, the beam configuration is non-redundant
in order to produce different spatial fringe periods, and thus to avoid crosstalk between the fringe peaks in the Fourier space. The separation $B_{\rm ij}$ between beams $i$ and $j$ in the output pupil is respectively equal to $3D$, $9D$ and $6D$, where $D$ is the beam diameter.
Since the thermal background at the longest wavelengths is variable and
much exceeds the target coherent flux, it is important to limit the crosstalk between the low frequency peak and the high frequency fringe peaks to a level below the thermal background photon noise limit. Two methods are used in MATISSE to ensure this result and estimate the coherent flux with high accuracy: spatial optical path difference (OPD) modulation, as in AMBER, and temporal OPD modulation, as in MIDI.
For each of the six baselines used, the
observable quantities, in each spectral channel, are the following:
\begin{itemize}
\item photometry;
\item coherent flux;
\item absolute visibility derived from the
­photometry and the coherent flux
measurements;
\item differential visibility (i.e.,
change of visibility with wavelength);
\item differential phase (i.e.,
change of phase with wavelength); 
\item closure phase.
\end{itemize}
To measure the visibility, we need to extract the source photometry by
separating the stellar flux from the sky background using sky chopping. However, the observation of the sky and the target, during chopping, are
never simultaneous. Therefore, thermal background fluctuations will be the most important contribution to the visibility error. Fortunately, chopping is unnecessary for measuring the coherent flux, and the differential and closure phases.
\begin{figure} [t]
   \begin{center}
   \begin{tabular}{c} %% tabular useful for creating an array of images 
   \includegraphics[scale=0.8]{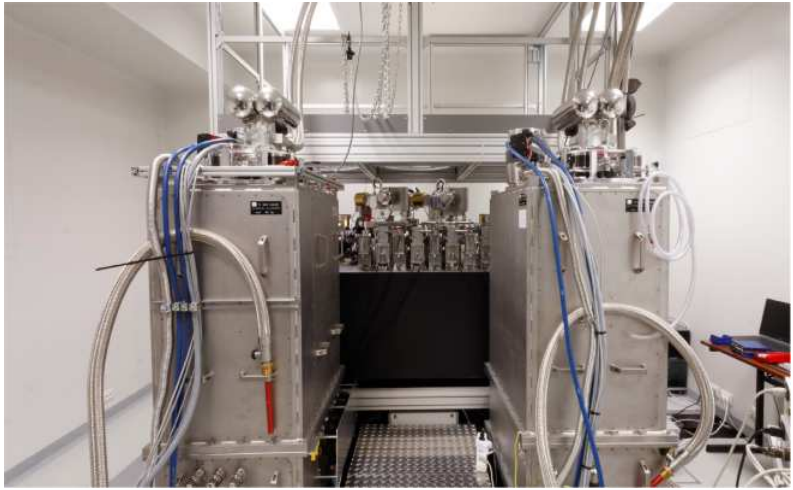}
   \end{tabular}
   \end{center}
   \caption[example] 
%>>>> use \label inside caption to get Fig. number with \ref{}
   { \label{fig:matissepicture} 
Current view of the MATISSE instrument, in the integration and tests room in Nice, with the two cryostats in the front and the warm optics table behind.}
   \end{figure} 
\section{Operating modes}
MATISSE has two standard operating modes. The HighSens mode does not
provide photometry and all photons are collected in the interferometric channel. This maximises the sen­sitivity for the coherent flux, the differential phase and the closure phases. In this mode, sequential photometric observations are nevertheless possible after the interferometric
observations. In the SiPhot mode, two thirds of the flux goes into the interferometric channel and one third into the photometric channels. Chopping is used to measure the average source photometry and extract the visibility from the coherent flux. These two modes can also
be mixed in a hybrid mode, which will combine for instance HighSens in {\itshape N} band and SiPhot in {\itshape L} band. \\
In the SiPhot mode, the detector simultaneously collects the light of up to five instrument outputs: the interferometric
­pattern plus four photometric signals (see Fig.~\ref{fig:detector}). During observations
with four telescopes, the interferogram
contains the combined dispersed fringe
pattern of six baselines. Each fringe pattern has a different spatial period due to the non-redundancy of the beam configuration. In the spatial direction, the minimum sampling of the interferometric signal requires 4 pixels per period of the narrowest fringes (24 pixels for the widest) at the short edge of the spectral bands. The sampling
of the interferometric beam is 72 pixel per $\lambda/D$ in the spatial direction and 3 pixels per $\lambda/D$ in the spectral direction, corresponding to an anamorphic factor of 24.
In the spatial direction, the interferometric field is about 468 pixels wide (corresponding to a field of $4\lambda/D$) and the photometric field is about 78 pixels. The size in the spectral direction depends on the spectral resolution and varies from 100 pixels for {\itshape L} and {\itshape M} bands at low spectral resolution (150 pixels for the
{\itshape N} band at low resolution) to the full detector for medium and high spectral resolution (indicated on Fig.~\ref{fig:detector}).

\section{Instrument design}
MATISSE is composed of the common Warm OPtics (WOP) and two Cold Optics
Benches (COB) that contain the two mid-infrared detectors, one covering the {\itshape L} and {\itshape M} bands, and one covering the {\itshape N} band). The two COBs are housed each in an independent cryostat (see Fig.~\ref{fig:matissevlti} where only a single COB is sketched). The location of the different parts of the instrument inside the VLTI laboratory is illustrated in Fig.~\ref{fig:matissevlti}. The WOP rests on a 2 by 1.5 meter optical table and receives four beams - designated as IP7/5/3/1 - through the feeding optics, coming from either the UTs or ATs. These four beams enter first into the beam commuting device, which allows the commutation of beams IP7 and IP5 and beams IP3 and IP1. The beams are then individually anamorphosed with a ratio of 1:4 by the cylindrical optics. The beams are then spectrally separated with individual dichroics in order to form the
 {\itshape L/M} and {\itshape N} band beams.\\ 
 Before entering into the cryostats, each beam passes through two modules: a periscope that is used for
 the co-alignment of the image and the pupil, and a delay line that delivers the pupil plane at the correct position
 into the COB. This delay line also equalizes the optical path difference between the beams and the differential optical path between the
 {\itshape L} and {\itshape M} bands and the {\itshape N} band. The WOP also contains the OPD modulation function, which is part of the spectral separator. In addition, the WOP accommodates two internal optical sources in a tower: a visible light source for alignment purposes (a fibered laser diode) and an infrared source for calibration purposes (a ceramic with thermal insulation
 housing). These internal optical sources deliver four identical beams and are injected into the instrument through the SOurce Selector module (SOS).\\
The cold optics benches consist of several modules and boxes. The beam selector cartridge holds four shutters. The re-imager box supports the cold stop in the pupil plane, several curved optics, and the spatial filters in
the image plane with its pinhole and slit slider. The beam-shaper box contains the beam splitters with a slider, several folding mirrors, the anamorphic optics and the photometric injection mirrors. The
wheel box includes several wheels to select the filtering, polarizing, and dispersive elements. Eventually, the camera box carries the two ­camera lenses, a folding mirror and the detector mount (see Fig~.\ref{fig:layout}).\\
Light passes the entrance windows of the cryostats with an anamorphic factor of 4. It passes the cold stops and the ­off-axis optics and spatial filtering module of the re-imager unit, until it reaches the beam splitter. At this stage, light is split into the interferometric channel and the photometric channels. The anamorphism of the interferometric channel is further increased to 24 by the anamorphic optics. Finally, after
passing the filter, polariser and dispersion wheels, light will reach the detector via the camera (see Fig.\ref{fig:layout}).\\
MATISSE uses two different detectors. The {\itshape L} and {\itshape M} bands detector is a Teledyne HAWAII-2RG of 2048 by 2048 pixels, grouped in 32 blocks of 64 by 2048 pixels. The {\itshape N} band detector is a Raytheon AQUARIUS, which has a format of 1024 by 1024 pixels, grouped in 2 times 32 blocks of 32 by 512 pixels.
\begin{figure}[t]
   \begin{center}
   \begin{tabular}{c} %% tabular useful for creating an array of images 
   \includegraphics[scale=0.5]{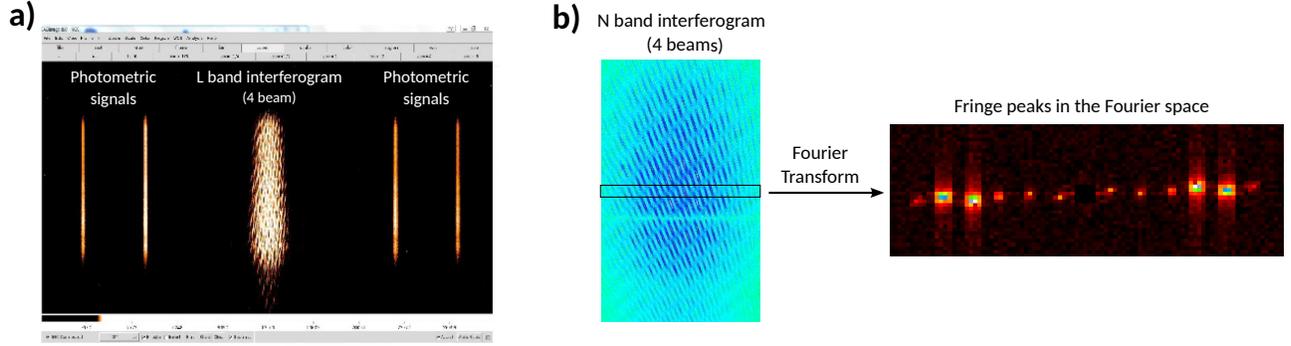}
   \end{tabular}
   \end{center}
   \caption[example] 
%>>>> use \label inside caption to get Fig. number with \ref{}
   { \label{fig:fringes} a) First lab fringes with 4 beams obtained in July 2015 on the HAWAII detector ({\itshape L} and {\itshape M} bands) with the IR artificial source, with two photometric signals on each side of the interferometric channel at the center; b) First lab {\itshape N} band fringes with 4 beams obtained in May 2016 on the AQUARIUS detector (left) with the IR artificial source, and the corresponding fringe peaks in the Fourier space (right).}
   \end{figure} 
\section{Performance}
Table \ref{tab:fluxes} gives the limiting correlated fluxes expected for MATISSE, with and without the use of an external fringe tracker. The values take into account the VLTI characteristics (e.g., optical transmission, adaptive optics performance, tip-tilt, focal laboratory) and a full calibration procedure \cite{2012SPIE.8445E..2JL}. The sensitivity performances with the fringe tracker corresponds to typical integration and observation times. These times are given as a guide and may thus be increased to reach higher sensitivities. The expected ultimate performance will require some evolution of the VLTI infrastructure: external fringe tracking, collecting data such as OPD and tip-tilt residuals, and lateral pupil motion monitoring or even active correction. The MATISSE specifications and expected performances, in particular with the use of an external fringe tracker, are summarized in Matter et al. (2016, this volume).

\begin{table}[h]
\caption{Expected sensitivity performances of MATISSE in SiPhot mode, without and with an external fringe tracker. Low spectral resolution ($R=30$) is considered. The integration and observation times with the fringe tracker are typical values, given as a guide, and may be increased to reach higher sensitivities.  
%Fonts sizes to be used for various parts of the manuscript.  Table captions should be centered above the table.  When the caption is too long to fit on one line, it should be justified to the right and left margins of the body of the text.
} 
\label{tab:fluxes}
\begin{center}       
\begin{tabular}{ccccc} %% this creates two columns
%% |l|l| to left justify each column entry
%% |c|c| to center each column entry
%% use of \rule[]{}{} below opens up each row
\hline
&\multicolumn{2}{c}{\textbf{{\itshape L} band sensitivity}} & \multicolumn{2}{c}{\textbf{{\itshape N} band sensitivity}} \\
\hline
&{\small Without FT} & {\small With FT}&{\small Without FT} &{\small With FT} \\
&&{\footnotesize (DIT=300ms)}&&{\footnotesize (Obs=10s)} \\
\hline
\textbf{AT} & {\small 2.95 Jy ({\itshape L}=5)} & {\small 0.55 Jy ({\itshape L}=6.8)}& {\small 14.6 Jy ({\itshape N}=1)} & {\small 2.1 Jy ({\itshape N}=3.1)} \\
\textbf{UT} & {\small 0.26 Jy ({\itshape L}=7.6)} & {\small 0.05 Jy ({\itshape L}=9.5)}& {\small 0.9 Jy ({\itshape N}=4)} & {\small 0.12 Jy ({\itshape N}=6.2)} \\
\hline 
\end{tabular}
\end{center}
\end{table}

\section{Status of the project}
\label{sec:status}
All the sub-systems of MATISSE are now fully integrated at the Observatoire de la Cote d'Azur (see Fig.~\ref{fig:matissepicture}). A first optical alignment of the WOP and the {\itshape L/M} band COB led to the first fringes with the artificial infrared source in July 2015 (see Fig.~\ref{fig:fringes}). However, humidity problems in the integration room caused damages to several of the MATISSE equipments and a 6-months delay in the initial planning. After interventions in the integration room and repair of the damaged equipments, the optical alignment phase could be restarted in early March this year. It is now finalized both in {\itshape L} and {\itshape M} bands, and in {\itshape N} band. This led in particular to the first fringes at 4 telescopes in {\itshape N} band (see Fig.~\ref{fig:fringes}).\\
The test phase started in June 2016
and will end in March 2017. As part of the current activities, some corrective actions were carried out to cancel some straylight observed on the {\itshape L/M} band detector. \\
Once the test phase has been completed, the ESO review called Preliminary
Acceptance Europe will be conducted between March 2017 and July 2017. It
will give the green light for the shipping of the instrument to Paranal
and its integration in September 2017. After the period called Assembly,
Integration and Verification, the Commissioning will take place in early
2018. If such a schedule is maintained, the first observing proposals can
be submitted for September 2018. This will lead to the first science in 2019.

\acknowledgments % equivalent to \section*{ACKNOWLEDGMENTS}       
MATISSE is defined, funded and built, in close collaboration with ESO, by a Consortium composed of French (INSU-CNRS in Paris and OCA in Nice), German (MPIA, MPIfR and University of Kiel), Dutch (NOVA and University of Leiden), and Austrian (University of Vienna) institutes. The Conseil G\'en\'eral des
Alpes-Maritimes in France, the Konkoly Observatory and the Cologne University also provided support to the instrument manufacturing.
We thank all MATISSE friends for their deep involvement and work and thus acknowledge C. Paladini, E. Thiebaut, K. Demick, Ph. Mathias, A. Niedzeilski, A. Russell, B. Stecklum, J. R. Walsh, A. C. da Fonte Martins, W. Boland, J.-M. Hameury, A. Crida, J. Colin, L. Pasquini, D. Mourard and A. Roussel for their assistance. Our special thoughts go to O. Chesneau, M. Dugue and
A. Moorwood, who left us too early.
 
% References
\bibliography{report} % bibliography data in report.bib
\bibliographystyle{spiebib} % makes bibtex use spiebib.bst

\end{document}